\documentclass[aps,prd,nofootinbib,amsmath,amssymb,superscriptaddress,onecolumn,11pt]{revtex4}

\pdfoutput=1

\usepackage{txfonts}
\usepackage{graphicx}
\usepackage{dcolumn}
\usepackage{bm}
\usepackage{amssymb}
\usepackage{latexsym}
\usepackage{booktabs}
\usepackage{amsmath}
\usepackage{multirow}
\usepackage[colorlinks=true, linkcolor=red, citecolor=blue]{hyperref}

\newcommand{\be}{\begin{equation}}
\newcommand{\ee}{\end{equation}}
\newcommand{\bq}{\begin{eqnarray}}
\newcommand{\eq}{\end{eqnarray}}

\begin{document}

\title{A search for sterile neutrinos with the latest cosmological observations}

\author{Lu Feng}
\affiliation{Department of Physics, College of Sciences, Northeastern University, Shenyang
110004, China}
\author{Jing-Fei Zhang}
\affiliation{Department of Physics, College of Sciences, Northeastern University, Shenyang
110004, China}
\author{Xin Zhang\footnote{Corresponding author}}
\email{zhangxin@mail.neu.edu.cn} \affiliation{Department of Physics, College of Sciences,
Northeastern University, Shenyang 110004, China}
\affiliation{Center for High Energy Physics, Peking University, Beijing 100080, China}

\begin{abstract}

We report the result of a search for sterile neutrinos with the latest cosmological observations. Both cases of massless and massive sterile neutrinos are considered in the $\Lambda$CDM cosmology. The cosmological observations used in this work include the Planck 2015 temperature and polarization data, the baryon acoustic oscillation data, the Hubble constant direct measurement data, the Planck Sunyaev-Zeldovich cluster counts data, the Planck lensing data, and the cosmic shear data. We find that the current observational data give a hint of the existence of massless sterile neutrino (as dark radiation) at the 1.44$\sigma$ level, and the consideration of an extra massless sterile neutrino can indeed relieve the tension between observations and improve the cosmological fit. For the case of massive sterile neutrino, the observations give a rather tight upper limit on the mass, which implies that actually a massless sterile neutrino is more favored. Our result is consistent with the recent result of neutrino oscillation experiment done by the Daya Bay and MINOS collaborations, as well as the recent result of cosmic ray experiment done by the IceCube collaboration.

\end{abstract}
\maketitle

\section{Introduction}
\label{sec1}
Since the discovery of the acceleration of expansion of the universe, dark energy has been proposed to explain the cause of the acceleration phenomenon \cite{Sahni:1999gb,Peebles:2002gy,Padmanabhan:2002ji,Frieman:2008sn,Weinberg:2012es,Mortonson:2013zfa,Wang:2016och}. The cosmological constant $\Lambda$ \cite{Einstein:1917ce} is the preferred candidate for dark energy.  In practice, the $\Lambda$ cold dark matter ($\Lambda$CDM) model has been achieving successes in fitting various observational data (it is still the best model so far in fitting data; see, e.g., Refs.~\cite{Li:2009jx,Xu:2016grp}). However, it has been found that several important astrophysical observations are inconsistent with the Planck observation (based on the $\Lambda$CDM cosmology), to some extent. For example, the fit results of the six-parameter base $\Lambda$CDM model based on the Planck+WP+highL data \cite{Ade:2013zuv} are in tension with the direct measurement of the Hubble constant \cite{Riess:2011yx}, the Planck Sunyaev-Zeldovich clusters counts \cite{Ade:2013lmv}, and the cosmic shear measurement of CFHTLenS survey \cite{Benjamin:2012qp}.

In our previous work \cite{Zhang:2014dxk}, we have shown that involving a light (sub-eV) sterile neutrino species in the $\Lambda$CDM model can help to reconcile the tensions between Planck and above-mentioned astrophysical observations (see also Refs.~\cite{Dvorkin:2014lea,Hamann:2013iba,Wyman:2013lza,Battye:2013xqa}). It was shown that \cite{Zhang:2014dxk}, under the CMB+BAO constraint, when taking the sterile neutrino into account, the inconsistencies with Planck are improved from 2.4$\sigma$ to 1.0$\sigma$ for $H_0$ observation, from 4.3$\sigma$ to 2.0$\sigma$ for SZ cluster counts observation, and from 2.3$\sigma$ to 1.7$\sigma$ for cosmic shear observation. For other relevant studies of sterile neutrinos, see, e.g., Refs.~\cite{Zhang:2014ifa,Zhang:2014nta,Ko:2014bka,Archidiacono:2014apa,Archidiacono:2014nda,Li:2014dja,An:2014bik,Zhang:2014lfa,Li:2015poa,Palazzo:2013me,deHolanda:2010am}. Thus, to relieve the tensions in cosmological observations, a natural choice is to consider the model with light sterile neutrinos.

The possibility of the existence of light sterile neutrinos has been motivated to explain the anomalies of short-baseline neutrino experiments \cite{Gariazzo:2013gua,Giunti:2013aea,Kopp:2013vaa,Giunti:2012bc,Giunti:2012tn,Aguilar-Arevalo:2012fmn,Conrad:2012qt,Mention:2011rk,Giunti:2010zu,Aguilar:2001ty}. It seems that the fully thermalized ($\Delta N_{\rm eff}\approx1$) sterile neutrinos with eV-scale mass are needed to explain these results \cite{Abazajian:2012ys,Hannestad:2012ky,Conrad:2013mka}. The cosmological observations play an important role in constraining the total mass of the active neutrinos (see, e.g., Refs.~\cite{Zhao:2016ecj,Wang:2016tsz,Huang:2015wrx,Zhang:2015uhk,Zhang:2015rha,Wang:2012uf,Lesgourgues:2006nd,Drewes:2013gca,GonzalezGarcia:2007ib,Zhang:2017rbg,Guo:2017hea,Capozzi:2017ipn,Li:2017iur,Vagnozzi:2017ovm,Lu:2016hsd,Capozzi:2016rtj,DellOro:2015kys,Giusarma:2016phn,DiValentino:2015ola,DiValentino:2016hlg,Yang:2017amu,Dai:2017sst,Zhu:2014qma}). In fact, since the sterile neutrinos have some effects on the evolution of the universe, the cosmological observations can also provide independent evidence in searching for sterile neutrinos \cite{Zhang:2014dxk}.

Recently, some important updated observational data were released. For example, Riess et al. \cite{Riess:2016jrr} reported the new result of local measurement of the Hubble constant, $H_0=73.00{\pm1.75}~{\rm km}~{\rm s}^{-1}~{\rm Mpc}^{-1}$, which is 3.3$\sigma$ higher than the result of  66.93$\pm$0.62$~{\rm km}~{\rm s}^{-1}~{\rm Mpc}^{-1}$ derived by the Planck mission based on the $\Lambda$CDM model with $\sum m_\nu=0.06$ eV using the latest Planck CMB data. Moreover, the BAO measurements were also updated for the CMASS and LOWZ galaxy samples, done by the Data Release 12 (DR12) \cite{Cuesta:2015mqa} of SDSS-III BOSS (Baryon Oscillation Spectroscopic Survey).

The aim of this work is to search for the sterile neutrinos by using the latest cosmological observations. We shall investigate how the sterile neutrinos can relieve the tensions in the current observations and whether the current cosmological data can provide evidence for the existence of sterile neutrino.

This paper is organized as follows. In Sec. \ref{sec2}, we introduce the analysis method and the observational data we use in this paper. The results are given and discussed in Sec. \ref{sec3}. Conclusion is given in Sec. \ref{sec4}.

\section{Analysis method and observational data}
\label{sec2}
In this section, we introduce the analysis method and the observational data that will be used in this paper.

\subsection{Analysis method}

We will consider the both cases of massless and massive sterile neutrinos in the framework of $\Lambda$CDM cosmology. The massless sterile neutrino serves as the dark radiation, and thus when this case is considered, an additional parameter $N_{\rm eff}$ needs to be added in the model; this case is called $\Lambda$CDM+$N_{\rm eff}$ model in this paper. When the sterile neutrino is considered to be massive, then one needs to add another extra parameter, $m_{\nu,{\rm sterile}}^{\rm eff}$, in the model; this case is thus called $\Lambda$CDM+$N_{\rm eff}$+$m_{\nu,{\rm sterile}}^{\rm eff}$ model in this paper.

In the case of massive sterile neutrino, the true mass of a thermally distributed sterile neutrino reads $m_{\rm sterile}^{\rm thermal}=(N_{\rm eff}-3.046)^{-3/4}m_{\nu,{\rm sterile}}^{\rm eff}$. In order to avoid a negative $m_{\rm sterile}^{\rm thermal}$, $N_{\rm eff}$ must be larger than 3.046 in a universe with sterile neutrinos. In this work, the active neutrino mass is kept at 0.06 eV (i.e., the minimal-mass normal hierarchy is assumed).

We place constraints on the $\Lambda$CDM cosmology with sterile neutrinos by using the current observational data. The conventions used in this paper are consistent with those adopted by the Planck collaboration \cite{Ade:2015xua}, i.e., those used in the {\tt camb} Boltzmann code \cite{Lewis:1999bs}.
There are six independent cosmological parameters in the base $\Lambda$CDM model,
$${\bf P}=\{\omega_b,~\omega_c,~100\theta_{\rm MC},~\tau,~\ln (10^{10}A_s),~n_s\},$$
where $\omega_b\equiv \Omega_b h^2$ and $\omega_c\equiv \Omega_c h^2$ are the present-day baryon and cold dark matter densities, respectively, $\theta_{\rm MC}$ is the ratio between the sound horizon and the angular diameter distance at the decoupling epoch, $\tau$ is the Thomson scattering optical depth due to reionization, $A_s$ is the amplitude of initial curvature perturbation power spectrum at $k=0.05~{\rm Mpc}^{-1}$, and $n_s$ is its  spectral index. In addition, there are two additional free parameters, $N_{\rm eff}$ and $m_{\nu,{\rm sterile}}^{\rm eff}$, for describing the sterile neutrino. Thus, there are seven independent parameters in total for the $\Lambda$CDM+$N_{\rm eff}$ model, and there are eight independent parameters in total for the $\Lambda$CDM+$N_{\rm eff}$+$m_{\nu,{\rm sterile}}^{\rm eff}$ model. Other parameters, such as $\Omega_{\rm m}$, $\sigma_8$, $H_0$, and so on, are the derived parameters.


We use the {\tt CosmoMC} package \cite{Lewis:2002ah} to infer the posterior probability distributions of parameters.
Flat priors for the base parameters are used. The prior ranges for the base parameters are chosen to be much wider than the posterior ranges in order not to affect the results of parameter estimation.

\subsection{Observational data}
In this paper, the data sets we use include the cosmic microwave background (CMB), the baryon acoustic oscillations (BAO), the Hubble constant ($H_0$), the Planck Sunyaev-Zeldovich (SZ), the Planck lensing, and the weak lensing (WL) observations.

{\it The CMB data}: We use the Planck 2015 CMB temperature and polarization data \cite{Aghanim:2015xee}
in our calculations. We consider the  combination of the likelihood at $30\leq \ell\leq2500$ in the temperature (TT), the cross-correlation of temperature and polarization (TE), and the polarization (EE) power spectra and the Planck low-$\ell$ likelihood in the range of $2\leq \ell\leq29$, which is denoted as ``Planck TT,TE,EE+lowP", following the nomenclature of the Planck collaboration \cite{Ade:2015xua}.

{\it The BAO data}: In order to break the geometric degeneracy, it is necessary to consider the BAO data. We use the LOWZ ($z_{\rm eff}=0.32$) and CMASS ($z_{\rm eff}=0.57$) samples of BOSS DR12 \cite{Cuesta:2015mqa},
as well as the 6dFGS (six-degree-field galaxy survey) ($z_{\rm eff}=0.106$) sample \cite{Beutler:2011hx}
and the SDSS MGS (main galaxy sample) ($z_{\rm eff}=0.15$) sample \cite{Ross:2014qpa}.

{\it The $H_0$ measurement}: We use the recently measured new local value of the Hubble constant, $H_0=73.00{\pm1.75}~{\rm km}~{\rm s}^{-1}~{\rm Mpc}^{-1}$, reported in Ref.~\cite{Riess:2016jrr}.

{\it The SZ data}: The counts of rich clusters of galaxies are from the sample of Planck thermal Sunyaev-Zeldovich (SZ) cluster observation \cite{Ade:2015fva}.

{\it The Lensing data}: We use the Planck lensing data \cite{Ade:2015zua}, which provide additional information at low redshift.

{\it The WL data}: We use the cosmic shear data of weak lensing (WL) from the CFHTLenS survey \cite{Heymans:2013fya}.

In what follows, we will use these observational data to place constraints on the $\Lambda$CDM cosmology with sterile neutrinos. We will compare the $\Lambda$CDM, $\Lambda$CDM+$N_{\rm eff}$, and $\Lambda$CDM+$N_{\rm eff}$+$m_{\nu,{\rm{sterile}}}^{\rm{eff}}$ models under the uniform data sets. The basic data combination adopted in this paper is the Planck TT,TE,EE+lowP+BAO combination. In order to show the impacts from the other astrophysical observations on measuring the properties of the sterile neutrino, we also further combine the $H_0$+SZ+Lensing+WL data in the analysis. Thus, in our analysis, we use the two data combinations: (i) Planck TT,TE,EE+lowP+BAO, and (ii) Planck TT,TE,EE+lowP+BAO+$H_0$+SZ+Lensing+WL. For convenience, we occasionally use the abbreviations ``CMB+BAO'' and ``CMB+BAO+other'' for them in the paper. In the next section, we will report and discuss the fitting results of the cosmological models in the light of these data sets.

\section{Results and discussion}\label{sec3}

\begin{table*}\small
\setlength\tabcolsep{1.0pt}
\renewcommand{\arraystretch}{1.0}
\caption{Fitting results for the $\Lambda$CDM model, the $\Lambda$CDM+$N_{\rm eff}$ model, and the $\Lambda$CDM+$N_{\rm eff}$+$m_{\nu,{\rm{sterile}}}^{\rm{eff}}$ model. We quote $\pm 1\sigma$ errors, but for the parameters that cannot be well constrained, we quote the 95.4\% CL upper limits.}
\centering
\begin{tabular}{ccccccccc}\hline
\hline Data &\multicolumn{3}{c}{Planck TT,TE,EE+lowP+BAO}&&\multicolumn{3}{c}{Planck TT,TE,EE+lowP+BAO+$H_{0}$+SZ+lensing+WL}&\\
           \cline{2-4}\cline{6-8}

  Model & $\Lambda$CDM &$\Lambda$CDM+$N_{\rm eff}$&$\Lambda$CDM+$N_{\rm eff}$+$m_{\nu,{\rm{sterile}}}^{\rm{eff}}$ &&$\Lambda$CDM &$\Lambda$CDM+$N_{\rm eff}$&$\Lambda$CDM+$N_{\rm eff}$+$m_{\nu,{\rm{sterile}}}^{\rm{eff}}$\\

\hline
$\Omega_bh^2$&$0.0223\pm0.0001$&$0.0225\pm0.0002$&$0.0224\pm0.0002$&&$0.0224\pm0.0001$&$0.0226\pm0.0002$&$0.0226\pm0.0002$\\
$\Omega_ch^2$&$0.1186\pm0.0010$&$0.1212^{+0.0017}_{-0.0027}$&$0.1187^{+0.0044}_{-0.0027}$&&$0.1175\pm0.0010$&$0.1214^{+0.0021}_{-0.0028}$&$0.1210^{+0.0026}_{-0.0031}$\\
$100\theta_{\rm MC}$&$1.0409\pm0.0003$&$1.0406\pm0.0004$&$1.0407^{+0.0004}_{-0.0003}$&&$1.0411\pm0.0003$&$1.0406\pm0.0004$&$1.0407\pm0.0004$\\
$\tau$&$0.085^{+0.017}_{-0.016}$&$0.088^{+0.018}_{-0.017}$&$0.090\pm0.017$&&$0.072\pm0.012$&$0.071\pm0.012$&$0.079\pm0.014$\\
${\rm{ln}}(10^{10}A_s)$&$3.103^{+0.033}_{-0.032}$&$3.114^{+0.036}_{-0.033}$&$3.116^{+0.034}_{-0.035}$&&$3.072\pm0.022$&$3.081\pm0.023$&$3.096\pm0.028$\\
$n_s$&$0.9676^{+0.0039}_{-0.0040}$&$0.9732^{+0.0055}_{-0.0064}$&$0.9719^{+0.0050}_{-0.0078}$&&$0.9699^{+0.0040}_{-0.0039}$&$0.9774^{+0.0056}_{-0.0058}$&$0.9789^{+0.0064}_{-0.0073}$\\
\hline
$m_{\nu,{\rm{sterile}}}^{\rm{eff}}$&...&...&$<0.7279$&&...&...&$<0.2417$\\
$N_{\rm eff}$&...&$<3.4384$&$<3.4273$&&...&$3.29^{+0.11}_{-0.17}$&$3.30^{+0.12}_{-0.20}$\\
\hline
$\Omega_m$&$0.3080^{+0.0061}_{-0.0062}$&$0.3048\pm0.0067$&$0.3085^{+0.0067}_{-0.0068}$&&$0.3011^{+0.0057}_{-0.0058}$&$0.2975^{+0.0059}_{-0.0064}$&$0.3010^{+0.0065}_{-0.0070}$\\
$\sigma_8$&$0.832\pm0.013$&$0.841\pm0.015$&$0.817^{+0.030}_{-0.021}$&&$0.8154^{+0.0085}_{-0.0084}$&$0.826^{+0.010}_{-0.011}$&$0.814^{+0.018}_{-0.013}$\\
$H_0$&$67.81^{+0.47}_{-0.46}$&$68.81^{+0.71}_{-1.02}$&$68.30^{+0.53}_{-1.00}$&&$68.33\pm0.45$&$69.71^{+0.84}_{-1.01}$&$69.44^{+0.84}_{-1.14}$\\
\hline
$\chi^2_{\rm min}$&12952.922&12952.976&12953.554&&13140.214&13138.042&13141.042\\
\hline
\hline
\end{tabular}
\label{tab1}
\end{table*}

\begin{figure*}[!htp]
\includegraphics[scale=1]{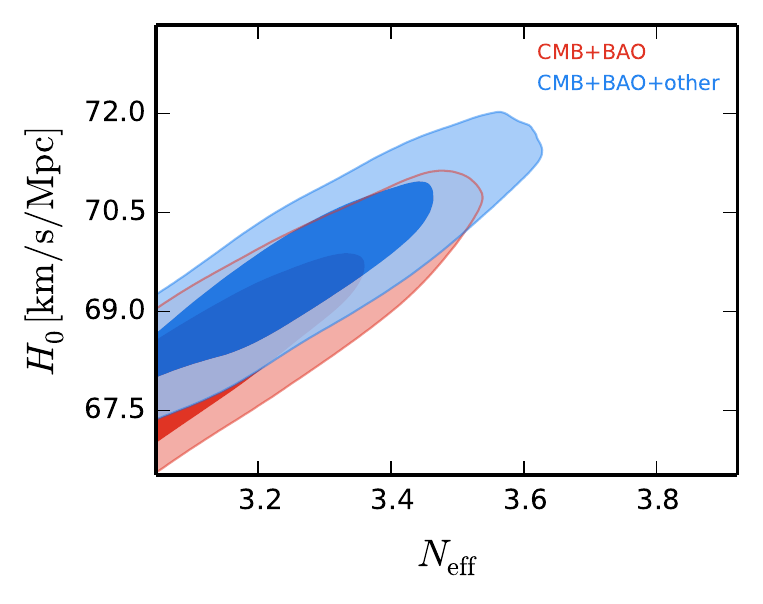}
\includegraphics[scale=0.9]{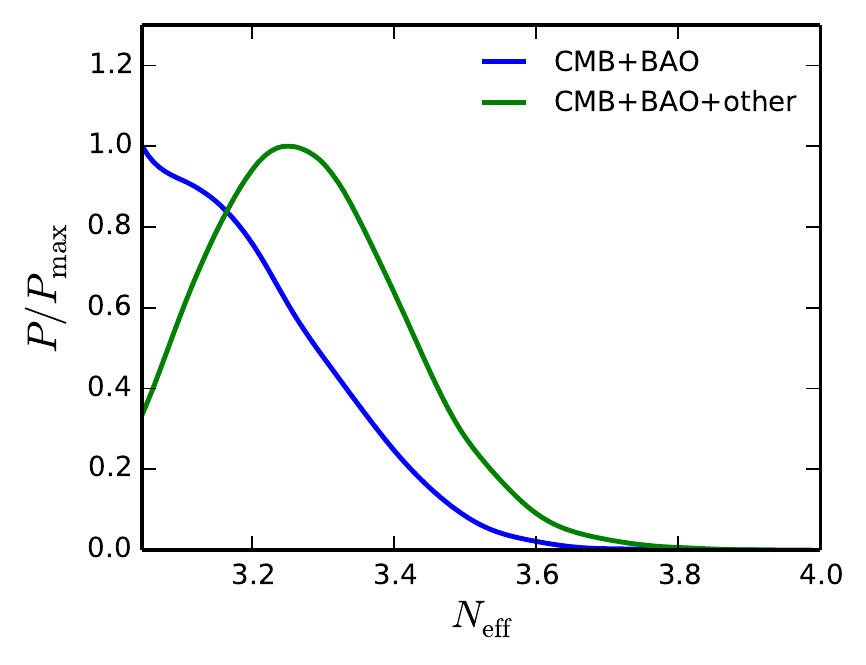}
\centering
\caption{\label{fig1}Constraint results for the $\Lambda$CDM+$N_{\rm eff}$ model from the CMB+BAO data combination and the CMB+BAO+other data combination. Left panel: two-dimensional marginalized posterior contours (68.3\% and 95.4\% CL) in the $N_{\rm eff}$--$H_0$ plane. Right panel: one-dimensional marginalized posterior distributions for $N_{\rm eff}$.}
\end{figure*}

\begin{figure*}[!htp]
\includegraphics[scale=1]{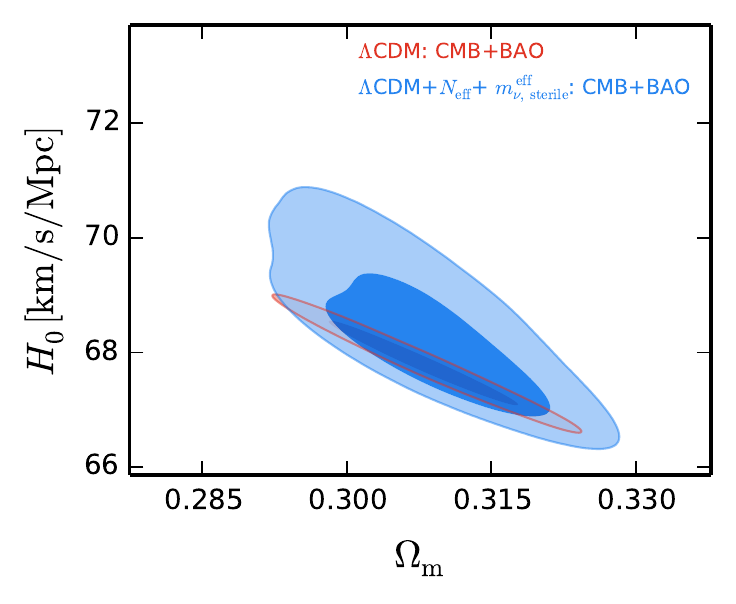}
\includegraphics[scale=1]{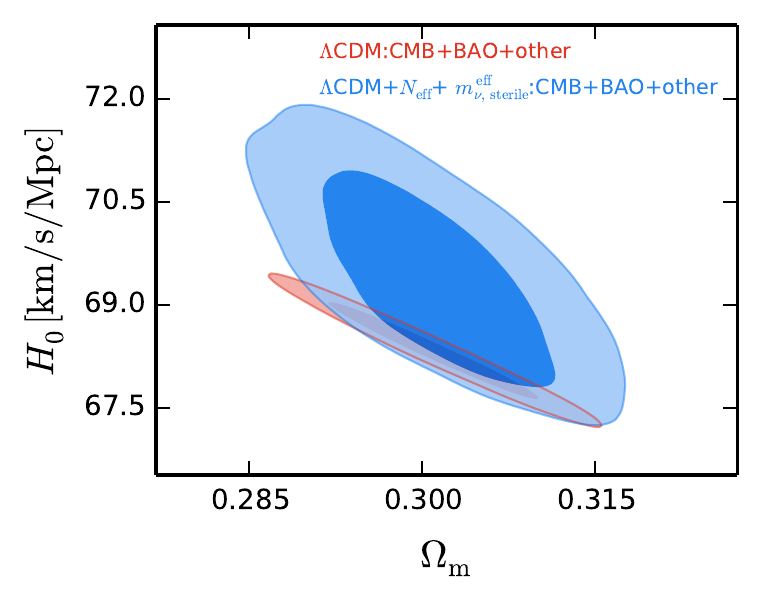}
\centering
 \caption{\label{fig2} The two-dimensional marginalized contours (68.3\% and 95.4\% CL) in the $\Omega_m$--$H_0$ plane, for the $\Lambda$CDM and $\Lambda$CDM+$N_{\rm eff}$+$m_{\nu,{\rm{sterile}}}^{\rm{eff}}$ models, from the constraints of CMB+BAO ({\it left}) and CMB+BAO+other ({\it right}).}
\end{figure*}

\begin{figure*}[!htp]
\includegraphics[scale=0.9]{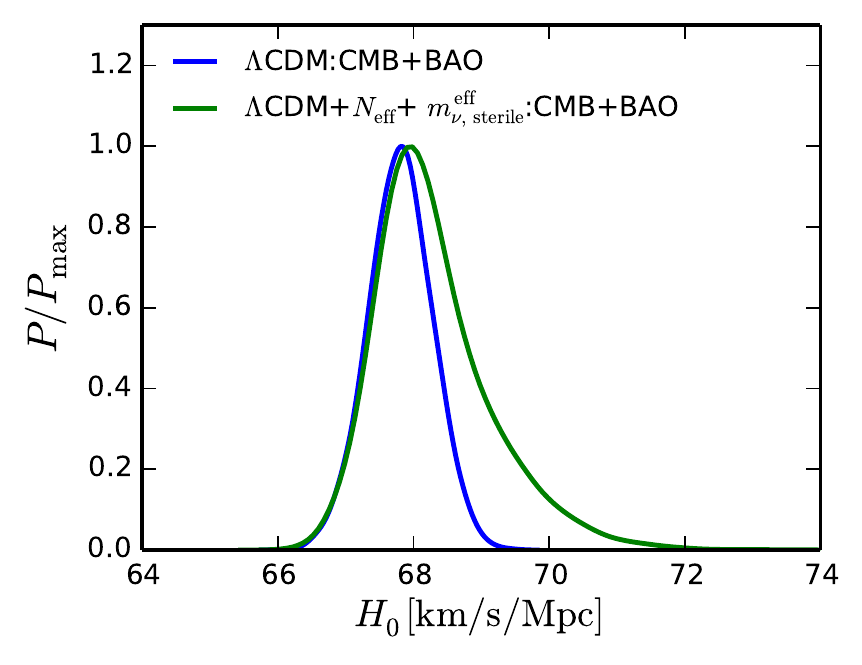}
\includegraphics[scale=0.9]{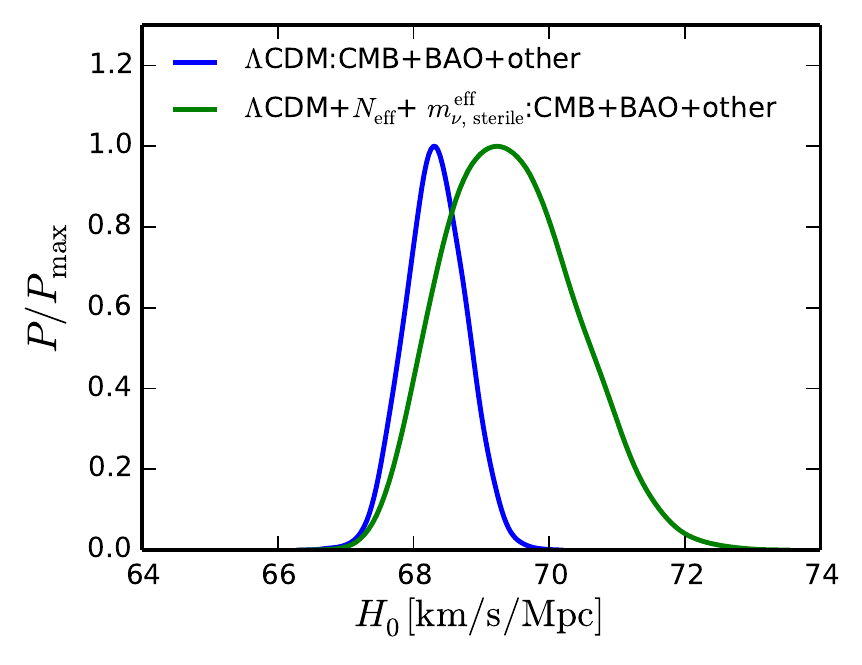}

\centering
 \caption{\label{fig3} The one-dimensional posterior distributions of $H_0$ in the $\Lambda$CDM and $\Lambda$CDM+$N_{\rm eff}$+$m_{\nu,{\rm{sterile}}}^{\rm{eff}}$ models, from the constraints of CMB+BAO ({\it left}) and CMB+BAO+other ({\it right}). }
\end{figure*}

\begin{figure*}[!htp]
\includegraphics[scale=0.3]{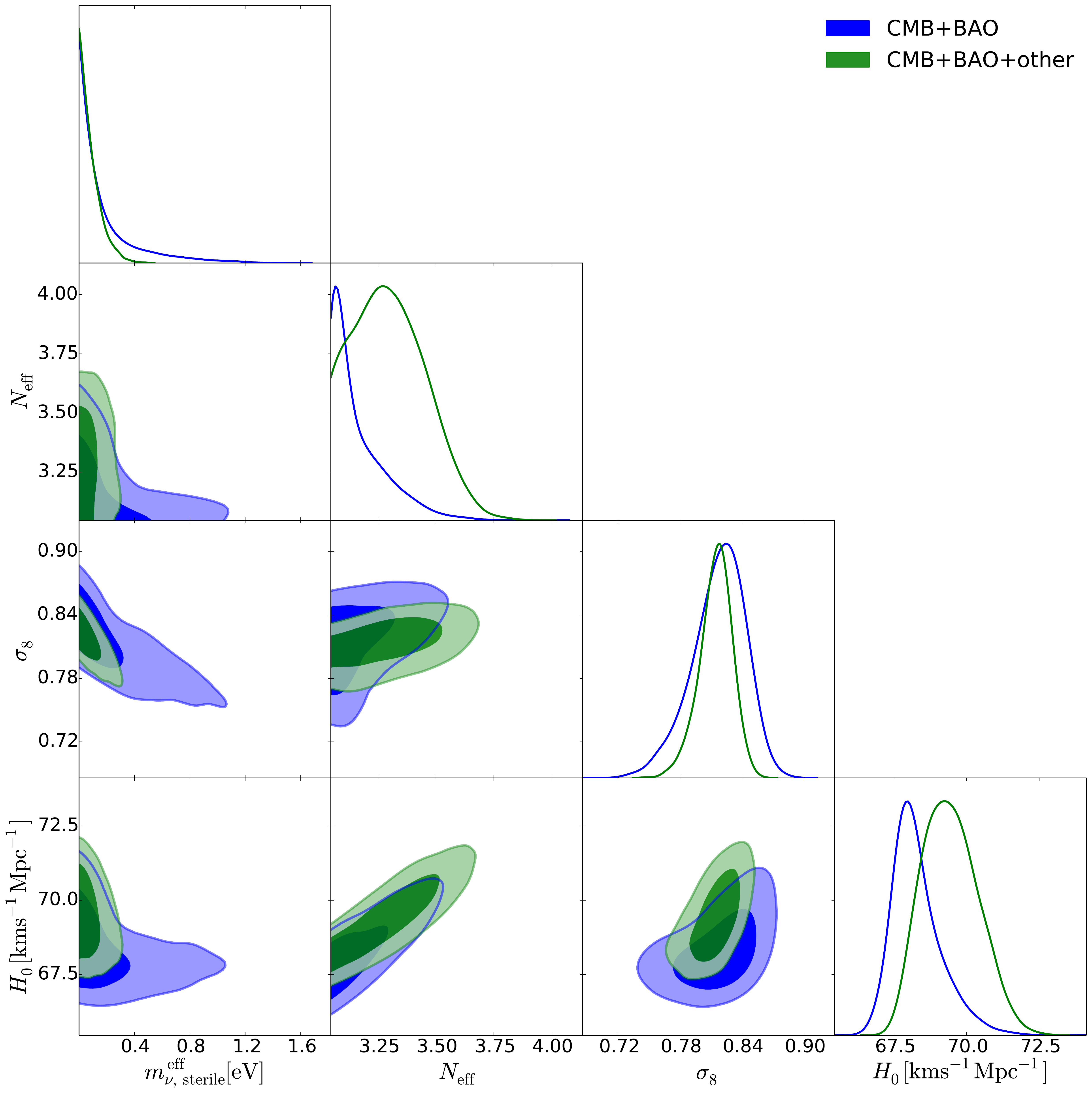}
\centering
 \caption{\label{fig4}The one-dimensional posterior distributions and two-dimensional marginalized contours (68.3\% and 95.4\% CL) for the $\Lambda$CDM+$N_{\rm eff}$+$m_{\nu,{\rm{sterile}}}^{\rm{eff}}$ model, from the constraints of the CMB+BAO and CMB+BAO+other data combinations.}
\end{figure*}


In this section, we report the fitting results of the cosmological models (the $\Lambda$CDM model, the $\Lambda$CDM+$N_{\rm eff}$ model, and the $\Lambda$CDM+$N_{\rm eff}$+$m_{\nu,{\rm sterile}}^{\rm eff}$ model) and discuss the implications of these results in the search for sterile neutrinos. We will discuss the cases of massless and massive sterile neutrinos, respectively, in the two subsections.

Detailed fit values for the three models for cosmological parameters are given in Table \ref{tab1}. In the table, we quote the $\pm1\sigma$ errors, but for the parameters that cannot be well constrained, we quote the 95.4\% CL upper limits.

When we make comparison for the three models from the statistical point of view, we must be aware of the fact that they have different numbers of parameters. In general, a model with more parameters tends to give a better fit to the same data, i.e., it tends to have a smaller $\chi_{\rm min}^2$. Thus, when the comparison is made for models with different parameter numbers, the simple comparison of $\chi_{\rm min}^2$ is not appropriate because it is unfair. A punishment mechanism must be considered for those models with more parameters. The simplest way, for our purpose in this work, is to consider the Akaike information criterion (AIC) \cite{AIC1974}, with the definition ${\rm AIC}=\chi_{\rm min}^2+2k$, where $k$ is the number of parameters of a model. A model with a lower AIC is more favored by data. So, when we make model selection for two models, we will calculate the difference of AIC for them, i.e., $\Delta {\rm AIC}=\Delta\chi_{\rm min}^2+2\Delta k$. In the case of this paper, we take the $\Lambda$CDM model as a reference model, and thus the $\Lambda$CDM+$N_{\rm eff}$ model has $\Delta k=1$ and the $\Lambda$CDM+$N_{\rm eff}$+$m_{\nu,{\rm sterile}}^{\rm eff}$ model has $\Delta k=2$. They are considered to be more favored over the $\Lambda$CDM model provided that the $\Lambda$CDM+$N_{\rm eff}$ model has $\Delta\chi^2<-2$ and the $\Lambda$CDM+$N_{\rm eff}$+$m_{\nu,{\rm sterile}}^{\rm eff}$ model has $\Delta\chi^2<-4$.

\subsection{The case of massless sterile neutrino}

The massless sterile neutrinos serve as the dark radiation, and thus in this case the effective number of relativistic species $N_{\rm eff}$ is treated as a free parameter. The total energy density of radiation in the universe is given by
$$\rho_r=[1+N_{\rm eff}\frac{7}{8}(\frac{4}{11})^{\frac{4}{3}}]\rho_\gamma,$$
where $\rho_\gamma$ is the energy density of photons. The standard case of three-generation neutrinos leads to $N_{\rm eff}=3.046$ \cite{Mangano:2001iu,Mangano:2005cc}. The detection of $\Delta N_{\rm eff}=N_{\rm eff}-3.046>0$ indicates the presence of extra relativistic particle species in the universe, and in this paper we take the fitting result of $\Delta N_{\rm eff}>0$ as the evidence of the existence of massless sterile neutrinos.

In this subsection, we constrain the $\Lambda$CDM+$N_{\rm eff}$ model by using two combinations of data sets, namely, the CMB+BAO and CMB+BAO+other combinations. As mentioned above, hereafter, we shall use ``CMB'' to denote Planck TT,TE,EE+lowP and use ``other" to denote $H_0$+SZ+lensing+WL. The free parameters include six base parameters and $N_{\rm eff}$ , and we fix the active neutrino mass $\sum m_\nu=0.06$ eV (two massless and one massive active neutrinos). The results of constraints on $N_{\rm eff}$ can be found in Table \ref{tab1}, and in this table the constraint results of other free parameters are also listed.

To directly show how the effective number of relativistic species $N_{\rm eff}$ affects the constraints on $H_0$,
we plot the two-dimensional posterior distribution contours (68.3\% and 95.4\% CL) in the $N_{\rm eff}$--$H_0$ plane for the $\Lambda$CDM+$N_{\rm eff}$ model, under the constraints from the two data combinations in the left panel of Fig.~\ref{fig1}. We find that, in the $\Lambda$CDM+$N_{\rm eff}$ model, $H_0$ is positively correlated with $N_{\rm eff}$, for the two data combinations. Moreover, under the constrains of CMB+BAO+other data, the $\Lambda$CDM model  gives $H_0=68.33{\pm0.45}~{\rm km}~{\rm s}^{-1}~{\rm Mpc}^{-1}$, and the $\Lambda$CDM+$N_{\rm eff}$ model gives $H_0=69.71^{+0.84}_{-1.01}~{\rm km}~{\rm s}^{-1}~{\rm Mpc}^{-1}$, indicating that the tension with the direct measurement of Hubble constant, $H_0=73.00{\pm1.75}~{\rm km}~{\rm s}^{-1}~{\rm Mpc}^{-1}$, is relieved to be at 2.58$\sigma$ level and 1.69$\sigma$ level, respectively. Therefore, the addition of the parameter $N_{\rm eff}$ can indeed relieve the tension between Planck data and $H_0$ direct measurement.

In the right panel of Fig.\ref{fig1}, we show the one-dimensional marginalized posterior distributions of $N_{\rm eff}$ for the $\Lambda$CDM+$N_{\rm eff}$ model using the two data combinations. We find that the CMB+BAO data can only give an upper limit, $N_{\rm eff}<3.44$ (95.4\% CL), but the CMB+BAO+other data can well constrain $N_{\rm eff}$, giving the result of $N_{\rm eff}=3.29^{+0.11}_{-0.17}$, which indicates a detection of $\Delta N_{\rm eff}>0$ at the 1.44$\sigma$ level.

Fitting to the CMB+BAO data, the $\Lambda$CDM and $\Lambda$CDM+$N_{\rm eff}$ models yield a similar $\chi^2_{\rm min}$, implying that with this data combination adding the parameter $N_{\rm eff}$ cannot improve the fit. But, when fitting to the CMB+BAO+other data, the $\Lambda$CDM+$N_{\rm eff}$ model leads to an increase of $\Delta\chi^2=-2.172$, compared to the $\Lambda$CDM model, indicating that in this case the addition of the parameter $N_{\rm eff}$ can evidently improve the fit. Actually, when we use the information criterion to make a model selection, we have $\Delta {\rm AIC}=-0.172$ for the $\Lambda$CDM+$N_{\rm eff}$ model in this case, which shows that the $\Lambda$CDM+$N_{\rm eff}$ model is only slightly better than the $\Lambda$CDM model from the statistical point of view.

Therefore, we find that the current CMB+BAO+other data can give a hint of the existence of massless sterile neutrino (as dark radiation) at the 1.44$\sigma$ level, and the consideration of an extra massless sterile neutrino can indeed relieve the tension between observations and improve the cosmological fit.

\subsection{The case of massive sterile neutrino}

In this subsection, we further consider the case of massive sterile neutrino, i.e., we consider two extra parameters, $N_{\rm eff}$ and $m_{\nu,{\rm sterile}}^{\rm eff}$, compared to the $\Lambda$CDM model.

We make a comparison for the $\Lambda$CDM model and $\Lambda$CDM+$N_{\rm eff}$+$m_{\nu,{\rm sterile}}^{\rm eff}$ model using the two data combinations. In Fig.~\ref{fig2}, we show the posterior distribution contours in the $\Omega_{\rm m}$--$H_0$ plane, for the two models. In the left panel, we show the case of fitting to the CMB+BAO data, and in the right panel, we show the case of fitting to the CMB+BAO+other data. We find that $H_0$ is anti-correlated with $\Omega_{\rm m}$ in the two models, and the consideration of massive sterile neutrino amplifies the parameter space, in particular, in the direction of $H_0$. Actually, the involvement of massive sterile neutrino can indeed give a much higher value of $H_0$.

We show the one-dimensional posterior distributions of $H_0$ for the two models in Fig.~\ref{fig3} (left panel for CMB+BAO and right panel for CMB+BAO+other). From this figure, we can clearly see that once the massive sterile neutrino is considered in cosmology, then under the constraint of CMB+BAO+other data the fit value of $H_0$ can become much larger. Under the constraints of CMB+BAO data, we obtain $H_0=67.81^{+0.47}_{-0.46}$ km s$^{-1}$ Mpc$^{-1}$ for the $\Lambda$CDM model and $H_0=68.30^{+0.53}_{-1.00}$ km s$^{-1}$ Mpc$^{-1}$ for the $\Lambda$CDM+$N_{\rm eff}$+$m_{\nu,{\rm sterile}}^{\rm eff}$ model, which indicates that the consideration of massive sterile neutrino improves the tension with the Hubble constant direct measurement ($H_0=73.00\pm 1.75$ km s$^{-1}$ Mpc$^{-1}$) from 2.86$\sigma$ to 2.57$\sigma$. So, in this case, the tension can only be slightly relieved. Under the constraints of CMB+BAO+other data, we obtain $H_0=68.33\pm 0.45$ km s$^{-1}$ Mpc$^{-1}$ for the $\Lambda$CDM model and $H_0=69.44^{+0.84}_{-1.14}$ km s$^{-1}$ Mpc$^{-1}$ for the $\Lambda$CDM+$N_{\rm eff}$+$m_{\nu,{\rm sterile}}^{\rm eff}$ model, which indicates that the involvement of massive sterile neutrino improves the tension from 2.58$\sigma$ to 1.83$\sigma$. Thus, in this case, the tension can be largely relieved. But we must admit that the improvement for the tension (from 2.58$\sigma$ to 1.83$\sigma$) is mainly owing to the fact that the introduction of two extra parameters leads to the posterior distribution of $H_0$ becoming much broader (see the right panel of Fig.~\ref{fig3}).

Then we show the constraint results for the parameters $N_{\rm eff}$ and $m_{\nu,{\rm sterile}}^{\rm eff}$ in the $\Lambda$CDM+$N_{\rm eff}$+$m_{\nu,{\rm sterile}}^{\rm eff}$ model (see Table \ref{tab1}):
$$
\left.
\begin{array}{c}
N_{\rm eff}< 3.4273 \\
m_{\nu,{\rm sterile}}^{\rm eff}< 0.7279~ {\rm eV}~
\end{array}
\right\} \quad\mbox{CMB+BAO},
$$
$$
\left.
\begin{array}{c}
N_{\rm eff}= 3.30^{+0.12}_{-0.20} \\
m_{\nu,{\rm sterile}}^{\rm eff}< 0.2417~ {\rm eV}~
\end{array}
\right\} \quad\mbox{CMB+BAO+other}.
$$
We find that $N_{\rm eff}$ cannot be well constrained using only the CMB+BAO data, but the addition of $H_0$, SZ, Lensing, and WL data can significantly improve the constraint on $N_{\rm eff}$, favoring $\Delta N_{\rm eff}>0$ at the 1.27$\sigma$ statistical significance. For the mass of sterile neutrino, the CMB+BAO data give $m_{\nu,{\rm sterile}}^{\rm eff}< 0.7279~ {\rm eV}$ ($95.4\%~{\rm CL}$), and further including the $H_0$+SZ+Lensing+WL data leads to the result of $m_{\nu,{\rm sterile}}^{\rm eff}< 0.2417~ {\rm eV}$ ($95.4\%~{\rm CL}$). Evidently, adding low-redshift data tightens the constraint on $m_{\nu,{\rm sterile}}^{\rm eff}$ significantly. This indicates that the SZ cluster data (as well as the $H_0$, Lensing, and WL data) play an important role in constraining the mass of sterile neutrino. In Fig.~\ref{fig4}, we show the one- and two-dimensional marginalized posterior distributions of the parameters $N_{\rm eff}$, $m_{\nu,{\rm sterile}}^{\rm eff}$, $\sigma_8$, and $H_0$, for the $\Lambda$CDM+$N_{\rm eff}$+$m_{\nu,{\rm sterile}}^{\rm eff}$ model. We find that, in the $\Lambda$CDM+$N_{\rm eff}$+$m_{\nu,{\rm sterile}}^{\rm eff}$ model, $\sigma_8$ is anti-correlated with $m_{\nu,{\rm sterile}}^{\rm eff}$, and $H_0$ is positively correlated with $N_{\rm eff}$, which leads to the fact that considering massive sterile neutrinos in cosmology can effectively relieve the tensions among the current observations.

Compared to the $\Lambda$CDM model, the $\Lambda$CDM+$N_{\rm eff}$+$m_{\nu,{\rm sterile}}^{\rm eff}$ model does not provide an improved fit to the current observational data. Considering the massive sterile neutrino in cosmology leads to an increase of $\Delta\chi^2=0.632$ under the CMB+BAO constraint and an increase of $\Delta\chi^2=0.828$ under the CMB+BAO+other constraint. That is to say, the $\Lambda$CDM+$N_{\rm eff}$+$m_{\nu,{\rm sterile}}^{\rm eff}$ model has $\Delta{\rm AIC}=4.632$ and 4.828 for the two data-combination cases, which shows that the massive sterile neutrino is not favored by the current cosmological observations.

Therefore, we find that the current CMB+BAO+other data give a rather tight upper limit on $m_{\nu,{\rm sterile}}^{\rm eff}$ and favor $\Delta N_{\rm eff}>0$ at the 1.27$\sigma$ level. Together with the constraint results for the massless sterile neutrino, we can conclude that the current observations do not seem to favor a massive sterile neutrino, but favor a massless sterile neutrino in some sense (only at the more than 1$\sigma$ statistical significance). Our result is consistent with the recent result of neutrino oscillation experiment by the Daya Bay and MINOS collaborations \cite{Adamson:2016jku}, as well as the recent result of cosmic ray experiment by the IceCube collaboration \cite{TheIceCube:2016oqi}.

\section{Conclusion}\label{sec4}

The aim of this work is to search for sterile neutrinos using the latest cosmological observations. We consider the two cases of massless and massive sterile neutrinos, corresponding to the $\Lambda$CDM+$N_{\rm eff}$ model and the $\Lambda$CDM+$N_{\rm eff}$+$m_{\nu,{\rm sterile}}^{\rm eff}$ model, respectively. The observational data used in this paper include the Planck TT,TE,EE+lowP data, the BAO data, the $H_0$ direct measurement, the Planck SZ cluster counts data, the Planck CMB lensing data, and the cosmic shear data.

For the $\Lambda$CDM+$N_{\rm eff}$ model, the CMB+BAO+other data give $N_{\rm eff}=3.29^{+0.11}_{-0.17}$ (68.3\% CL), favoring $\Delta N_{\rm eff}=N_{\rm eff}-3.046>0$ at the 1.44$\sigma$ level. Therefore, there is a hint of the existence of massless sterile neutrinos, from the current cosmological observations. We also find that the addition of the parameter $N_{\rm eff}$ can indeed relieve the tension between the Planck observation and the recent $H_0$ direct measurement (the tension is reduced to be at the 1.69$\sigma$ level).

For the $\Lambda$CDM+$N_{\rm eff}$+$m_{\nu,{\rm sterile}}^{\rm eff}$ model, using the CMB+BAO data, we obtain $m_{\nu,{\rm sterile}}^{\rm eff}<0.7279$ eV (95.4$\%$ CL) and $N_{\rm eff}<3.4273$ (95.4\% CL). Thus, in this case, only upper limits on $N_{\rm eff}$ and $m_{\nu,{\rm sterile}}^{\rm eff}$ (for the massive sterile neutrino) can be derived. Further including the other ($H_0$+SZ+Lensing+WL) data significantly improves the constraints, and in this case we obtain $m_{\nu,{\rm sterile}}^{\rm eff}<0.2417$ eV (95.4\% CL) and $N_{\rm eff}=3.3^{+0.12}_{-0.20}$ (68.3\% CL). Thus, the current observations give a rather tight upper limit on $m_{\nu,{\rm sterile}}^{\rm eff}$ and favor $\Delta N_{\rm eff}>0$ at the 1.27$\sigma$ level. This result seems to favor a massless sterile neutrino, in tension with the previous short-baseline neutrino oscillation experiments that prefer the mass of sterile neutrino at around 1~eV. But our result is consistent with the recent result of neutrino oscillation experiment done by the Daya Bay and MINOS collaborations \cite{Adamson:2016jku}, as well as the recent result of cosmic ray experiment done by the IceCube collaboration \cite{TheIceCube:2016oqi}.

\begin{acknowledgments}
This work was supported by the National Natural Science Foundation of China (Grants No.~11522540 and No.~11690021), the Top-Notch Young Talents Program of China, and the Provincial Department of Education of Liaoning (Grant No.~L2012087).

\end{acknowledgments}


\begin{thebibliography}{99}

\bibitem{Sahni:1999gb}
  V.~Sahni and A.~A.~Starobinsky,
  The case for a positive cosmological lambda-term,
  Int.\ J.\ Mod.\ Phys.\ D {\bf 9}, 373 (2000)
  [astro-ph/9904398].
\bibitem{Peebles:2002gy}
  P.~J.~E.~Peebles and B.~Ratra,
  The Cosmological constant and dark energy,
  Rev.\ Mod.\ Phys.\  {\bf 75}, 559 (2003)
  [astro-ph/0207347].
\bibitem{Padmanabhan:2002ji}
  T.~Padmanabhan,
  Cosmological constant: the weight of the vacuum,
  Phys.\ Rep.\  {\bf 380}, 235 (2003)
  [hep-th/0212290].
\bibitem{Frieman:2008sn}
  J.~Frieman, M.~Turner and D.~Huterer,
  Dark Energy and the Accelerating Universe,
  Ann.\ Rev.\ Astron.\ Astrophys.\  {\bf 46}, 385 (2008)
  [arXiv:0803.0982 [astro-ph]].
\bibitem{Weinberg:2012es}
  D.~H.~Weinberg, M.~J.~Mortonson, D.~J.~Eisenstein, C.~Hirata, A.~G.~Riess and E.~Rozo,
  Observational probes of cosmic acceleration,
  Phys.\ Rep.\  {\bf 530}, 87 (2013)
  [arXiv:1201.2434 [astro-ph.CO]].
\bibitem{Mortonson:2013zfa}
  M.~J.~Mortonson, D.~H.~Weinberg and M.~White,
  Dark energy: a short review,
  [arXiv:1401.0046 [astro-ph.CO]].

\bibitem{Wang:2016och}
  S.~Wang, Y.~Wang and M.~Li,
  Holographic Dark Energy,
  arXiv:1612.00345 [astro-ph.CO].


\bibitem{Einstein:1917ce}
  A.~Einstein,
  Cosmological Considerations in the General Theory of Relativity,
  Sitzungsber.\ Preuss.\ Akad.\ Wiss.\ Berlin (Math.\ Phys.\ ) {\bf 1917}, 142 (1917).

\bibitem{Li:2009jx}
  M.~Li, X.~Li and X.~Zhang,
  Comparison of dark energy models: A perspective from the latest observational data,
  Sci.\ China Phys.\ Mech.\ Astron.\  {\bf 53}, 1631 (2010)
  [arXiv:0912.3988 [astro-ph.CO]].

\bibitem{Xu:2016grp}
  Y.~Y.~Xu and X.~Zhang,
  Comparison of dark energy models after Planck 2015,
  Eur.\ Phys.\ J.\ C {\bf 76}, no. 11, 588 (2016)
  [arXiv:1607.06262 [astro-ph.CO]].

\bibitem{Ade:2013zuv}
  P.~A.~R.~Ade {\it et al.} [Planck Collaboration],
  Planck 2013 results. XVI. Cosmological parameters,
  Astron.\ Astrophys.\  {\bf 571}, A16 (2014)
  [arXiv:1303.5076 [astro-ph.CO]].
\bibitem{Riess:2011yx}
  A.~G.~Riess {\it et al.},
  A 3$\% $Solution: Determination of the Hubble Constant with the Hubble Space Telescope and Wide Field Camera 3,
  Astrophys.\ J.\  {\bf 730}, 119 (2011)
  Erratum: [Astrophys.\ J.\  {\bf 732}, 129 (2011)]
  [arXiv:1103.2976 [astro-ph.CO]].
\bibitem{Ade:2013lmv}
  P.~A.~R.~Ade {\it et al.} [Planck Collaboration],
  Planck 2013 results. XX. Cosmology from Sunyaev¨CZeldovich cluster counts,
  Astron.\ Astrophys.\  {\bf 571}, A20 (2014)
  [arXiv:1303.5080 [astro-ph.CO]].
\bibitem{Benjamin:2012qp}
  J.~Benjamin {\it et al.},
  CFHTLenS tomographic weak lensing: Quantifying accurate redshift distributions,
  Mon.\ Not.\ Roy.\ Astron.\ Soc.\  {\bf 431}, 1547 (2013)
  [arXiv:1212.3327 [astro-ph.CO]].

\bibitem{Zhang:2014dxk}
  J.~F.~Zhang, Y.~H.~Li and X.~Zhang,
  Sterile neutrinos help reconcile the observational results of primordial gravitational waves from Planck and BICEP2,
  Phys.\ Lett.\ B {\bf 740}, 359 (2015)
  [arXiv:1403.7028 [astro-ph.CO]].


\bibitem{Dvorkin:2014lea}
  C.~Dvorkin, M.~Wyman, D.~H.~Rudd and W.~Hu,
  Neutrinos help reconcile Planck measurements with both the early and local Universe,
  Phys.\ Rev.\ D {\bf 90}, no. 8, 083503 (2014)
  [arXiv:1403.8049 [astro-ph.CO]].

\bibitem{Hamann:2013iba}
  J.~Hamann and J.~Hasenkamp,
  A new life for sterile neutrinos: resolving inconsistencies using hot dark matter,
  JCAP {\bf 1310}, 044 (2013)
  [arXiv:1308.3255 [astro-ph.CO]].
\bibitem{Wyman:2013lza}
  M.~Wyman, D.~H.~Rudd, R.~A.~Vanderveld and W.~Hu,
  Neutrinos Help Reconcile Planck Measurements with the Local Universe,
  Phys.\ Rev.\ Lett.\  {\bf 112}, no. 5, 051302 (2014)
  [arXiv:1307.7715 [astro-ph.CO]].
\bibitem{Battye:2013xqa}
  R.~A.~Battye and A.~Moss,
  Evidence for Massive Neutrinos from Cosmic Microwave Background and Lensing Observations,
  Phys.\ Rev.\ Lett.\  {\bf 112}, no. 5, 051303 (2014)
  [arXiv:1308.5870 [astro-ph.CO]].

\bibitem{Palazzo:2013me}
  A.~Palazzo,
  Phenomenology of light sterile neutrinos: a brief review,
  Mod.\ Phys.\ Lett.\ A {\bf 28}, 1330004 (2013)
  [arXiv:1302.1102 [hep-ph]].
\bibitem{Ko:2014bka}
  P.~Ko and Y.~Tang,
  $\nu\Lambda$MDM: A model for sterile neutrino and dark matter reconciles cosmological and neutrino oscillation data after BICEP2,
  Phys.\ Lett.\ B {\bf 739}, 62 (2014)
  [arXiv:1404.0236 [hep-ph]].
\bibitem{Archidiacono:2014apa}
  M.~Archidiacono, N.~Fornengo, S.~Gariazzo, C.~Giunti, S.~Hannestad and M.~Laveder,
  Light sterile neutrinos after BICEP-2,
  JCAP {\bf 1406}, 031 (2014)
  [arXiv:1404.1794 [astro-ph.CO]].
\bibitem{Zhang:2014nta}
  J.~F.~Zhang, Y.~H.~Li and X.~Zhang,
  Cosmological constraints on neutrinos after BICEP2,
  Eur.\ Phys.\ J.\ C {\bf 74}, 2954 (2014)
  [arXiv:1404.3598 [astro-ph.CO]].
\bibitem{Archidiacono:2014nda}
  M.~Archidiacono, S.~Hannestad, R.~S.~Hansen and T.~Tram,
  Cosmology with self-interacting sterile neutrinos and dark matter - A pseudoscalar model,
  Phys.\ Rev.\ D {\bf 91}, no. 6, 065021 (2015)
  [arXiv:1404.5915 [astro-ph.CO]].
\bibitem{Li:2014dja}
  Y.~H.~Li, J.~F.~Zhang and X.~Zhang,
  Tilt of primordial gravitational wave spectrum in a universe with sterile neutrinos,
  Sci.\ China Phys.\ Mech.\ Astron.\  {\bf 57}, 1455 (2014)
  [arXiv:1405.0570 [astro-ph.CO]].
\bibitem{An:2014bik}
  F.~P.~An {\it et al.} [Daya Bay Collaboration],
  Search for a Light Sterile Neutrino at Daya Bay,
  Phys.\ Rev.\ Lett.\  {\bf 113}, 141802 (2014)
  [arXiv:1407.7259 [hep-ex]].
\bibitem{Zhang:2014ifa}
  J.~F.~Zhang, J.~J.~Geng and X.~Zhang,
  Neutrinos and dark energy after Planck and BICEP2: data consistency tests and cosmological parameter constraints,
  JCAP {\bf 1410}, no. 10, 044 (2014)
  [arXiv:1408.0481 [astro-ph.CO]].
\bibitem{Zhang:2014lfa}
  J.~F.~Zhang, Y.~H.~Li and X.~Zhang,
  Measuring growth index in a universe with sterile neutrinos,
  Phys.\ Lett.\ B {\bf 739}, 102 (2014)
  [arXiv:1408.4603 [astro-ph.CO]].
\bibitem{Li:2015poa}
  Y.~H.~Li, J.~F.~Zhang and X.~Zhang,
  Probing $f(R)$ cosmology with sterile neutrinos via measurements of scale-dependent growth rate of structure,
  Phys.\ Lett.\ B {\bf 744}, 213 (2015)
  [arXiv:1502.01136 [astro-ph.CO]].
\bibitem{deHolanda:2010am}
  P.~C.~de Holanda and A.~Y.~Smirnov,
  Solar neutrino spectrum, sterile neutrinos and additional radiation in the Universe,
  Phys.\ Rev.\ D {\bf 83}, 113011 (2011)
  [arXiv:1012.5627 [hep-ph]].



\bibitem{Gariazzo:2013gua}
  S.~Gariazzo, C.~Giunti and M.~Laveder,
  Light Sterile Neutrinos in Cosmology and Short-Baseline Oscillation Experiments,
  JHEP {\bf 1311}, 211 (2013)
  [arXiv:1309.3192 [hep-ph]].
\bibitem{Giunti:2013aea}
  C.~Giunti, M.~Laveder, Y.~F.~Li and H.~W.~Long,
  Pragmatic View of Short-Baseline Neutrino Oscillations,
  Phys.\ Rev.\ D {\bf 88}, 073008 (2013)
  [arXiv:1308.5288 [hep-ph]].
\bibitem{Kopp:2013vaa}
  J.~Kopp, P.~A.~N.~Machado, M.~Maltoni and T.~Schwetz,
  Sterile Neutrino Oscillations: The Global Picture,
  JHEP {\bf 1305}, 050 (2013)
  [arXiv:1303.3011 [hep-ph]].
\bibitem{Giunti:2012bc}
  C.~Giunti, M.~Laveder, Y.~F.~Li and H.~W.~Long,
  Short-baseline electron neutrino oscillation length after troitsk,
  Phys.\ Rev.\ D {\bf 87}, no. 1, 013004 (2013)
  [arXiv:1212.3805 [hep-ph]].
\bibitem{Giunti:2012tn}
  C.~Giunti, M.~Laveder, Y.~F.~Li, Q.~Y.~Liu and H.~W.~Long,
  Update of Short-Baseline Electron Neutrino and Antineutrino Disappearance,
  Phys.\ Rev.\ D {\bf 86}, 113014 (2012)
  [arXiv:1210.5715 [hep-ph]].
\bibitem{Aguilar-Arevalo:2012fmn}
  A.~A.~Aguilar-Arevalo {\it et al.} [MiniBooNE Collaboration],
  A Combined $\nu_\mu \rightarrow \nu_e$ and $\bar \nu_\mu \rightarrow \bar \nu_e$ Oscillation Analysis of the MiniBooNE Excesses,
  arXiv:1207.4809 [hep-ex].
\bibitem{Conrad:2012qt}
  J.~M.~Conrad, C.~M.~Ignarra, G.~Karagiorgi, M.~H.~Shaevitz and J.~Spitz,
  Sterile Neutrino Fits to Short Baseline Neutrino Oscillation Measurements,
  Adv.\ High Energy Phys.\  {\bf 2013}, 163897 (2013)
  [arXiv:1207.4765 [hep-ex]].
\bibitem{Mention:2011rk}
  G.~Mention, M.~Fechner, T.~Lasserre, T.~A.~Mueller, D.~Lhuillier, M.~Cribier and A.~Letourneau,
  The Reactor Antineutrino Anomaly,
  Phys.\ Rev.\ D {\bf 83}, 073006 (2011)
  [arXiv:1101.2755 [hep-ex]].
\bibitem{Giunti:2010zu}
  C.~Giunti and M.~Laveder,
  Statistical Significance of the Gallium Anomaly,
  Phys.\ Rev.\ C {\bf 83}, 065504 (2011)
  [arXiv:1006.3244 [hep-ph]].
\bibitem{Aguilar:2001ty}
  A.~Aguilar-Arevalo {\it et al.} [LSND Collaboration],
  Evidence for neutrino oscillations from the observation of anti-neutrino(electron) appearance in a anti-neutrino(muon) beam,
  Phys.\ Rev.\ D {\bf 64}, 112007 (2001)
  [hep-ex/0104049].
\bibitem{Abazajian:2012ys}
  K.~N.~Abazajian {\it et al.},
  Light Sterile Neutrinos: A White Paper,
  arXiv:1204.5379 [hep-ph].
\bibitem{Hannestad:2012ky}
  S.~Hannestad, I.~Tamborra and T.~Tram,
  Thermalisation of light sterile neutrinos in the early universe,
  JCAP {\bf 1207}, 025 (2012)
  [arXiv:1204.5861 [astro-ph.CO]].
\bibitem{Conrad:2013mka}
  J.~M.~Conrad, W.~C.~Louis and M.~H.~Shaevitz,
  The LSND and MiniBooNE Oscillation Searches at High $\Delta m^2$,
  Ann.\ Rev.\ Nucl.\ Part.\ Sci.\  {\bf 63}, 45 (2013)
  [arXiv:1306.6494 [hep-ex]].

\bibitem{Zhao:2016ecj}
  M.~M.~Zhao, Y.~H.~Li, J.~F.~Zhang and X.~Zhang,
  Constraining neutrino mass and extra relativistic degrees of freedom in dynamical dark energy models using Planck 2015 data in combination with low-redshift cosmological probes: basic extensions to $\Lambda$CDM cosmology,
 Mon.\ Not.\ Roy.\ Astron.\ Soc.\  {\bf 469}, 1713 (2017)
  [arXiv:1608.01219 [astro-ph.CO]].


\bibitem{Wang:2016tsz}
  S.~Wang, Y.~F.~Wang, D.~M.~Xia and X.~Zhang,
  Impacts of dark energy on weighing neutrinos: mass hierarchies considered,
  Phys.\ Rev.\ D {\bf 94}, no. 8, 083519 (2016)
  [arXiv:1608.00672 [astro-ph.CO]].
\bibitem{Huang:2015wrx}
  Q.~G.~Huang, K.~Wang and S.~Wang,
  Constraints on the neutrino mass and mass hierarchy from cosmological observations,
  Eur.\ Phys.\ J.\ C {\bf 76}, no. 9, 489 (2016)
  [arXiv:1512.05899 [astro-ph.CO]].
\bibitem{Zhang:2015uhk}
  X.~Zhang,
  Impacts of dark energy on weighing neutrinos after Planck 2015,
  Phys.\ Rev.\ D {\bf 93}, no. 8, 083011 (2016)
  [arXiv:1511.02651 [astro-ph.CO]].
\bibitem{Zhang:2015rha}
  J.~F.~Zhang, M.~M.~Zhao, Y.~H.~Li and X.~Zhang,
  Neutrinos in the holographic dark energy model: constraints from latest measurements of expansion history and growth of structure,
  JCAP {\bf 1504}, 038 (2015)
  [arXiv:1502.04028 [astro-ph.CO]].
\bibitem{Wang:2012uf}
  Y.~H.~Li, S.~Wang, X.~D.~Li and X.~Zhang,
  Holographic dark energy in a Universe with spatial curvature and massive neutrinos: a full Markov Chain Monte Carlo exploration,
  JCAP {\bf 1302}, 033 (2013)
  [arXiv:1207.6679 [astro-ph.CO]].


\bibitem{Lesgourgues:2006nd}
  J.~Lesgourgues and S.~Pastor,
  Massive neutrinos and cosmology,
  Phys.\ Rept.\  {\bf 429}, 307 (2006)
  [astro-ph/0603494].
\bibitem{Drewes:2013gca}
  M.~Drewes,
  The Phenomenology of Right Handed Neutrinos,
  Int.\ J.\ Mod.\ Phys.\ E {\bf 22}, 1330019 (2013)
  [arXiv:1303.6912 [hep-ph]].

\bibitem{GonzalezGarcia:2007ib}
  M.~C.~Gonzalez-Garcia and M.~Maltoni,
  Phenomenology with Massive Neutrinos,
  Phys.\ Rept.\  {\bf 460}, 1 (2008)
  [arXiv:0704.1800 [hep-ph]].


\bibitem{Zhang:2017rbg}
  X.~Zhang,
  Weighing neutrinos in dynamical dark energy models,
  Sci.\ China Phys.\ Mech.\ Astron.\  {\bf 60}, no. 6, 060431 (2017)
  [arXiv:1703.00651 [astro-ph.CO]].

\bibitem{Guo:2017hea}
  R.~Y.~Guo, Y.~H.~Li, J.~F.~Zhang and X.~Zhang,
  Weighing neutrinos in the scenario of vacuum energy interacting with cold dark matter: application of the parameterized post-Friedmann approach,
  JCAP {\bf 1705}, no. 05, 040 (2017)
  [arXiv:1702.04189 [astro-ph.CO]].


\bibitem{Capozzi:2017ipn}
  F.~Capozzi, E.~Di Valentino, E.~Lisi, A.~Marrone, A.~Melchiorri and A.~Palazzo,
  Global constraints on absolute neutrino masses and their ordering,
  arXiv:1703.04471 [hep-ph].

\bibitem{Li:2017iur}
  E.~K.~Li, H.~Zhang, M.~Du, Z.~H.~Zhou and L.~Xu,
  Probing the Neutrino Mass Hierarchy with Dynamical Dark Energy Model,
  arXiv:1703.01554 [astro-ph.CO].


\bibitem{Vagnozzi:2017ovm}
  S.~Vagnozzi, E.~Giusarma, O.~Mena, K.~Freese, M.~Gerbino, S.~Ho and M.~Lattanzi,
  Unveiling $\nu$ secrets with cosmological data: neutrino masses and mass hierarchy,
  arXiv:1701.08172 [astro-ph.CO].

\bibitem{Lu:2016hsd}
  J.~Lu, M.~Liu, Y.~Wu, Y.~Wang and W.~Yang,
  Cosmic constraint on massive neutrinos in viable $f(R)$ gravity with producing $\Lambda$CDM background expansion,
  Eur.\ Phys.\ J.\ C {\bf 76}, no. 12, 679 (2016)
  [arXiv:1606.02987 [astro-ph.CO]].

\bibitem{Capozzi:2016rtj}
  F.~Capozzi, E.~Lisi, A.~Marrone, D.~Montanino and A.~Palazzo,
  Neutrino masses and mixings: Status of known and unknown $3\nu$ parameters,
  Nucl.\ Phys.\ B {\bf 908}, 218 (2016)
  [arXiv:1601.07777 [hep-ph]].


\bibitem{DellOro:2015kys}
  S.~Dell'Oro, S.~Marcocci, M.~Viel and F.~Vissani,
  The contribution of light Majorana neutrinos to neutrinoless double beta decay and cosmology,
  JCAP {\bf 1512}, no. 12, 023 (2015)
  [arXiv:1505.02722 [hep-ph]].

\bibitem{Giusarma:2016phn}
  E.~Giusarma, M.~Gerbino, O.~Mena, S.~Vagnozzi, S.~Ho and K.~Freese,
  Improvement of cosmological neutrino mass bounds,
  Phys.\ Rev.\ D {\bf 94}, no. 8, 083522 (2016)
  [arXiv:1605.04320 [astro-ph.CO]].


\bibitem{DiValentino:2015ola}
  E.~Di Valentino, A.~Melchiorri and J.~Silk,
  Beyond six parameters: extending $\Lambda$CDM,
  Phys.\ Rev.\ D {\bf 92}, no. 12, 121302 (2015)
  [arXiv:1507.06646 [astro-ph.CO]].

\bibitem{DiValentino:2016hlg}
  E.~Di Valentino, A.~Melchiorri and J.~Silk,
  Reconciling Planck with the local value of $H_0$ in extended
parameter space,
  arXiv:1606.00634 [astro-ph.CO].

\bibitem{Yang:2017amu}
  W.~Yang, R.~C.~Nunes, S.~Pan and D.~F.~Mota,
  Effects of neutrino mass hierarchies on dynamical dark energy models,
  arXiv:1703.02556 [astro-ph.CO].

\bibitem{Dai:2017sst}
  W.~M.~Dai, Z.~K.~Guo, R.~G.~Cai and Y.~Z.~Zhang,
  Lorentz invariance violation in the neutrino sector: a joint analysis from big bang nucleosynthesis and the cosmic microwave background,
  arXiv:1701.02553 [astro-ph.CO].

\bibitem{Zhu:2014qma}
  H.~M.~Zhu, U.~L.~Pen, X.~Chen and D.~Inman,
  Probing Neutrino Hierarchy and Chirality via Wakes,
  Phys.\ Rev.\ Lett.\  {\bf 116}, no. 14, 141301 (2016)
  [arXiv:1412.1660 [astro-ph.CO]].

\bibitem{Riess:2016jrr}
  A.~G.~Riess {\it et al.},
  A 2.4$\%$ Determination of the Local Value of the Hubble Constant,
  Astrophys.\ J.\  {\bf 826}, no. 1, 56 (2016)
  [arXiv:1604.01424 [astro-ph.CO]].
\bibitem{Cuesta:2015mqa}
  A.~J.~Cuesta {\it et al.},
  The clustering of galaxies in the SDSS-III Baryon Oscillation Spectroscopic Survey: Baryon Acoustic Oscillations in the correlation function of LOWZ and CMASS galaxies in Data Release 12,
  Mon.\ Not.\ Roy.\ Astron.\ Soc.\  {\bf 457}, no. 2, 1770 (2016)
  [arXiv:1509.06371 [astro-ph.CO]].

\bibitem{Ade:2015xua}
  P.~A.~R.~Ade {\it et al.} [Planck Collaboration],
  Planck 2015 results. XIII. Cosmological parameters,
  arXiv:1502.01589 [astro-ph.CO].
\bibitem{Lewis:1999bs}
  A.~Lewis, A.~Challinor and A.~Lasenby,
  Efficient computation of CMB anisotropies in closed FRW models,
  Astrophys.\ J.\  {\bf 538}, 473 (2000)
  [astro-ph/9911177].

\bibitem{AIC1974}
Akaike H. A new look at the statistical model identification. IEEE Trans Automatic Control, 1974, 19:716-723

\bibitem{Lewis:2002ah}
  A.~Lewis and S.~Bridle,
  Cosmological parameters from CMB and other data: A Monte Carlo approach,
  Phys.\ Rev.\ D {\bf 66}, 103511 (2002)
  [astro-ph/0205436].



\bibitem{Aghanim:2015xee}
  N.~Aghanim {\it et al.} [Planck Collaboration],
 Planck 2015 results. XI. CMB power spectra, likelihoods, and robustness of parameters,
  [arXiv:1507.02704 [astro-ph.CO]].


\bibitem{Beutler:2011hx}
  F.~Beutler {\it et al.},
  The 6dF Galaxy Survey: Baryon Acoustic Oscillations and the Local Hubble Constant,
  Mon.\ Not.\ Roy.\ Astron.\ Soc.\  {\bf 416}, 3017 (2011)
  [arXiv:1106.3366 [astro-ph.CO]].
\bibitem{Ross:2014qpa}
  A.~J.~Ross, L.~Samushia, C.~Howlett, W.~J.~Percival, A.~Burden and M.~Manera,
  The clustering of the SDSS DR7 main Galaxy sample ¨C I. A 4 per cent distance measure at $z = 0.15$,
  Mon.\ Not.\ Roy.\ Astron.\ Soc.\  {\bf 449}, no. 1, 835 (2015)
  [arXiv:1409.3242 [astro-ph.CO]]



\bibitem{Ade:2015fva}
  P.~A.~R.~Ade {\it et al.} [Planck Collaboration],
  Planck 2015 results. XXIV. Cosmology from Sunyaev-Zeldovich cluster counts,
  arXiv:1502.01597 [astro-ph.CO].

\bibitem{Ade:2015zua}
  P.~A.~R.~Ade {\it et al.} [Planck Collaboration],
  Planck 2015 results. XV. Gravitational lensing,
  arXiv:1502.01591 [astro-ph.CO].

\bibitem{Heymans:2013fya}
  C.~Heymans {\it et al.},
  CFHTLenS tomographic weak lensing cosmological parameter constraints: Mitigating the impact of intrinsic galaxy alignments,
  Mon.\ Not.\ Roy.\ Astron.\ Soc.\  {\bf 432}, 2433 (2013)
  [arXiv:1303.1808 [astro-ph.CO]].

\bibitem{Mangano:2001iu}
  G.~Mangano, G.~Miele, S.~Pastor and M.~Peloso,
  A Precision calculation of the effective number of cosmological neutrinos,
  Phys.\ Lett.\ B {\bf 534}, 8 (2002)
  [astro-ph/0111408].
\bibitem{Mangano:2005cc}
  G.~Mangano, G.~Miele, S.~Pastor, T.~Pinto, O.~Pisanti and P.~D.~Serpico,
  Relic neutrino decoupling including flavor oscillations,
  Nucl.\ Phys.\ B {\bf 729}, 221 (2005)
  [hep-ph/0506164].


\bibitem{Adamson:2016jku}
  P.~Adamson {\it et al.} [Daya Bay and MINOS Collaborations],
  Limits on Active to Sterile Neutrino Oscillations from Disappearance Searches in the MINOS, Daya Bay, and Bugey-3 Experiments,
  Phys.\ Rev.\ Lett.\  {\bf 117}, no. 15, 151801 (2016)
  Addendum: [Phys.\ Rev.\ Lett.\  {\bf 117}, no. 20, 209901 (2016)]
  [arXiv:1607.01177 [hep-ex]].

\bibitem{TheIceCube:2016oqi}
  M.~G.~Aartsen {\it et al.} [IceCube Collaboration],
  Searches for Sterile Neutrinos with the IceCube Detector,
  Phys.\ Rev.\ Lett.\  {\bf 117}, no. 7, 071801 (2016)
  [arXiv:1605.01990 [hep-ex]].









\end{thebibliography}
\end{document}